\pdfoutput=1
\documentclass[11pt,a4paper]{article}

\usepackage[utf8]{inputenc}
\usepackage[T1]{fontenc}
\usepackage[english]{babel}
\usepackage{geometry}
\usepackage{hyperref}
\usepackage{enumitem}
\usepackage{booktabs}
\usepackage{array}
\usepackage{xcolor}
\usepackage{listings}

\hypersetup{
  colorlinks=true,
  linkcolor=blue,
  citecolor=blue,
  urlcolor=blue
}

\lstdefinestyle{code}{
  basicstyle=\ttfamily\small,
  breaklines=true,
  frame=single,
  columns=fullflexible,
  keepspaces=true,
  showstringspaces=false
}
\title{\textbf{ESAA-Conversational:}\\
An Event-Sourced Memory Layer for Continuity, Handoff, and Curation Across Heterogeneous LLM Coding Agents}

\author{
Elzo Brito dos Santos Filho\\
\texttt{elzo.santos@cps.sp.gov.br}
}

\begin{document}

\maketitle

\begin{abstract}
Software developers increasingly work with multiple LLM coding agents, switching among tools such as Codex, Grok, Claude Code, and other assistants as context windows fill, sessions end, or a particular agent becomes better suited to a subtask. Each agent, however, persists its conversation in a private and vendor-specific log. The result is conversational state drift: goals, decisions, open tasks, and rationales established with one agent are not reliably available when another agent takes over. This paper presents \emph{ESAA-Conversational}, a domain specialization of Event-Sourcing Agent Architecture (ESAA)~\cite{esaa} for shared conversational memory across heterogeneous agents. The method treats the visible conversation as a local event store: hooks and watchers capture visible turns, normalize them into an append-only \texttt{activity.jsonl}, and deterministically project read models such as \texttt{handoff.md}, \texttt{state.md}, \texttt{decisions.md}, and \texttt{tasks.json}. Mechanical capture does not require LLM inference; agents use judgment only for explicit curation, recording durable decisions and conversational tasks through domain commands. The public v1.1.0 release implements a PowerShell CLI with \texttt{init}, \texttt{enable-hooks}, \texttt{sync}, \texttt{project}, \texttt{verify}, \texttt{context}, \texttt{decide}, and \texttt{task}; includes \texttt{workspace\_root} isolation and a write-path lockfile; and is distributed as a greenfield package with an empty public log. A self-referential case study with 570 development-lab events shows that heterogeneous agents can collaborate through a shared log without a direct agent-to-agent channel, while the public distribution preserves privacy by excluding the private conversational history.
\end{abstract}

\noindent\textbf{Keywords:} autonomous agents; event sourcing; conversational memory; agent handoff; context engineering; multi-agent systems; LLMs; operational auditability.

\section{Introduction}

LLM-based software engineering has moved beyond an isolated interaction with a single assistant. A developer may use Codex for local editing, Grok for architectural review, Claude Code for context reading, and other tools for specialized analysis. This poly-agentic practice is natural: agents differ in quality, cost, integration surface, and context-window limits. The problem is that continuity across those agents still tends to depend on the human.

When an agent session ends, or when the user switches tools, the next agent rarely knows the full conversational history. It may not know which alternatives were rejected, which decisions have already been made, which tasks remain open, or what operational state is expected. The usual remedy is to copy and paste context. This mechanism is manual, lossy, and expensive in tokens. It also conflates two activities that should remain separate: capturing conversational evidence and interpreting that evidence.

ESAA-Conversational proposes to treat continuity as an event-sourcing problem~\cite{fowler-event-sourcing}. Visible conversation is captured mechanically into an append-only log; useful state for agents is projected deterministically; durable decisions and tasks are recorded explicitly through curation commands. The system therefore governs shared memory among agents, not the engineering work itself.

The central distinction is:

\begin{lstlisting}
esaa-core
  governs agentic work over a project

conversation-esaa
  governs memory, continuity, handoff,
  curation, and synchronization across agents
\end{lstlisting}

This paper revises the initial formulation of ESAA-Conversational in light of the public v1.1.0 release. The revision incorporates decisions made during design and publication: a greenfield tool, no manual editing of read models, \texttt{activity.jsonl} as the single source of truth, paginated context commands, \texttt{enable-hooks} for operational setup, workspace isolation, and a clear separation between implemented v1.1 capabilities and future features such as snapshots, cold replay, forensic auditability, and advanced redaction.

\subsection{Contributions}

This paper makes five contributions:

\begin{enumerate}[leftmargin=*]
  \item It defines ESAA-Conversational as a domain specialization of ESAA for shared conversational memory across heterogeneous agents.
  \item It formalizes inverted ingestion: instead of requiring every agent to speak a common protocol, the system reads native logs, hooks, or watchers and normalizes visible turns into events.
  \item It establishes a strong boundary between mechanical capture and curation: turns are evidence, while durable decisions and tasks are explicit events produced by agent judgment.
  \item It describes the public v1.1.0 release, including a domain CLI, projected read models, paginated context, \texttt{workspace\_root} isolation, and a write-path lockfile.
  \item It reports a self-referential case study in which the system itself was designed and reviewed by distinct agents using the conversational memory being built.
\end{enumerate}

\section{Background and Related Work}

This work builds on three prior artifacts in the ESAA family: ESAA as the
event-sourced architecture for governing long-horizon agentic work~\cite{esaa},
PARCER as an output-governance contract for individual agent responses~\cite{parcer},
and ESAA-Security as an example of specializing the ESAA kernel to a concrete
domain~\cite{esaa-security}. It also adopts the broader software architecture
ideas of Event Sourcing~\cite{fowler-event-sourcing} and CQRS~\cite{fowler-cqrs},
and is motivated in part by empirical work on degradation in long-context
language-model use~\cite{lost-middle}.

\subsection{The ESAA Family}

The original ESAA approach establishes that long-horizon agentic work should be governed by events: the immutable log is the source of truth, while derived states are verifiable projections~\cite{esaa}. Instead of trusting a transient agent state or the mutable repository snapshot, the system records intentions, decisions, effects, and verification in an append-only trail.

ESAA-Conversational preserves this core, but changes the governed object. ESAA core governs what an agent does to a project; ESAA-Conversational governs what agents know about the conversation. This shift follows the broader ESAA pattern of specializing the event vocabulary to a domain, as also explored in ESAA-Security~\cite{esaa-security}. Conversation events, curated decisions, conversational tasks, and context windows are not equivalent to issues, roadmaps, or implementation tasks in ESAA core.

\subsection{Agent Memory}

Recent approaches to agent memory often emphasize semantic retrieval, embeddings, vector databases, generated wikis, or persistent summaries. These techniques are useful, but they address a different layer of the problem. ESAA-Conversational does not begin as semantic memory; it begins as a deterministic ledger. The primary question is not ``which passage seems relevant?'', but rather ``which event occurred, who produced it, in which workspace, and which read model can be reconstructed from it?''.

This choice does not exclude semantic search. Embeddings, generated wikis, and richer indexes may be added later as derived projections over the event store. The architectural contribution is to prevent those indexes from becoming the source of truth.

\subsection{Handoff and Context Engineering}

Attention degradation in long contexts motivates compact views rather than full transcripts~\cite{lost-middle}. ESAA-Conversational applies this idea to handoff across agents: a cold agent should begin with \texttt{handoff.md}, \texttt{state.md}, \texttt{decisions.md}, \texttt{tasks.json}, and a selective window obtained through \texttt{context}; it should not consume the entire \texttt{activity.jsonl} by default.

\section{Domain Model}

The ESAA-Conversational domain is composed of five main concepts:

\begin{description}[leftmargin=!,labelwidth=3.2cm]
  \item[Turn] Chronological visible evidence from the conversation.
  \item[Decision] Durable curated knowledge, with rationale and sources.
  \item[Task] Conversational operational continuity.
  \item[Handoff] An entry contract for a cold agent.
  \item[Window] A paginated, deterministic, on-demand slice of the event store.
\end{description}

The central rule is:

\begin{lstlisting}
turns     = evidence
decisions = durable knowledge
tasks     = operational continuity
handoff   = operational entry point
context   = selective log reading
\end{lstlisting}

In v1.1, snapshots, advanced redaction, forensic auditability, and full cold replay are positioned as future work. The v1.1 release prioritizes a simple local tool: PowerShell, files, deterministic tests, and a single CLI.

\section{Architecture}

\subsection{Overview}

The logical architecture separates the write path, the event store, and the read path, following the same broad separation of command-side mutation and query-side projection used in CQRS~\cite{fowler-cqrs}:

\begin{lstlisting}
agent log / hook / watcher
  -> conversation-esaa CLI
  -> lock
  -> append activity.jsonl
  -> project read models
  -> verify
  -> handoff/context for the next agent
\end{lstlisting}

The event store is \texttt{.conversation-esaa/activity.jsonl}. It is append-only in the public workflow. Files such as \texttt{state.md}, \texttt{handoff.md}, \texttt{decisions.md}, and \texttt{tasks.json} are reconstructible read models; they are not edited manually.

\subsection{Inverted Ingestion}

In ESAA core, the agent emits structured intentions to the runtime. That model works when the agent is inside the ESAA protocol. Commercial heterogeneous agents, however, do not share a common runtime. They expose logs, hooks, configuration files, stop events, or, in the worst case, surfaces that can be observed by polling.

ESAA-Conversational inverts the direction:

\begin{lstlisting}
native agent log
  -> deterministic adapter
  -> conversation_turn
  -> activity.jsonl
\end{lstlisting}

The hook does not implement domain logic. It calls the CLI:

\begin{lstlisting}[language=bash]
conversation-esaa sync --agent grok --workspace C:\project
conversation-esaa sync --agent claude --workspace C:\project
conversation-esaa sync --agent codex --workspace C:\project
\end{lstlisting}

This decision keeps hooks thin, auditable, and replaceable. Deduplication, normalization, projection, and verification belong to the local runtime.

\subsection{Source of Truth and Read Models}

Table~\ref{tab:artifacts} summarizes the main artifacts.

\begin{table}[h]
\centering
\begin{tabular}{p{4.1cm}p{3.0cm}p{6.4cm}}
\toprule
\textbf{Artifact} & \textbf{Type} & \textbf{Role} \\
\midrule
\texttt{activity.jsonl} & source of truth & Append-only log of turns, decisions, and conversational tasks. \\
\texttt{sync-state.json} & derived cache & Deduplication state; reconstructible from the log. \\
\texttt{state.md} & read model & Compact summary of conversational state. \\
\texttt{handoff.md} & read model & Entry contract for a cold agent. \\
\texttt{decisions.md} & read model & Projected list of curated decisions. \\
\texttt{tasks.json} & read model & Conversational backlog projected from \texttt{task.*} events. \\
\texttt{run/*.lock} & operational control & Lockfile for serializing the write path. \\
\bottomrule
\end{tabular}
\caption{ESAA-Conversational v1.1 artifacts.}
\label{tab:artifacts}
\end{table}

Two points are critical. First, \texttt{tasks.json} is not an editable surface; it is born as a projection. Second, \texttt{decisions.md} is not editable either; decisions enter through \texttt{conversation-esaa decide}. This removes dual writes and preserves the event-sourcing contract.

\subsection{Event Model}

The primary mechanical event is \texttt{conversation\_turn}. A v1.1 event must carry an identifier, timestamp, event type, actor, agent, and \texttt{workspace\_root}.

\begin{lstlisting}
{
  "event_id": "evt_...",
  "ts": "2026-06-21T13:44:18-03:00",
  "event": "conversation_turn",
  "actor": "assistant",
  "agent_id": "codex",
  "workspace_root": "C:\\xampp\\htdocs\\project",
  "summary": "Short turn summary",
  "text": "Visible turn text"
}
\end{lstlisting}

Curated events have their own vocabulary:

\begin{lstlisting}
decision.recorded
task.created
task.updated
task.closed
\end{lstlisting}

An example decision event is:

\begin{lstlisting}
{
  "event": "decision.recorded",
  "actor": "assistant",
  "agent_id": "codex",
  "workspace_root": "C:\\xampp\\htdocs\\project",
  "decision": "Agents read context through paginated commands",
  "rationale": "Avoid loading the full activity.jsonl",
  "related_turns": ["evt_001", "evt_018"]
}
\end{lstlisting}

This distinction resolves an important tension: the system does not merely capture verbatim turns; it also provides a native surface for durable knowledge.

\subsection{Workspace Isolation}

A common failure mode in local synchronizers is mixing logs from different projects. The v1.1 release adopts \texttt{workspace\_root} as a mandatory boundary for new events. The \texttt{context} command filters by workspace before applying filters by agent, topic, or window. Legacy events without this field may be read only in laboratory compatibility mode; the public package is greenfield and does not depend on them.

\subsection{Concurrency}

The v1.1 release uses a lockfile around the write pipeline:

\begin{lstlisting}
sync / decide / task
  -> acquire lock
  -> append event(s)
  -> project
  -> verify
  -> release lock
\end{lstlisting}

A single-writer queue, hash chain, or external anchoring would be stronger alternatives, but they are deferred to future versions. The decision is deliberate: v1.1 prioritizes a local, simple, and verifiable tool.

\section{Domain CLI}

The public v1.1.0 release exposes a single PowerShell CLI:

\begin{lstlisting}
conversation-esaa init --workspace <path>
conversation-esaa enable-hooks --agent <grok|claude|codex> --workspace <path> [--trust] [--watcher]
conversation-esaa sync --agent <grok|claude|codex> --workspace <path>
conversation-esaa project --workspace <path>
conversation-esaa verify --workspace <path>
conversation-esaa context --workspace <path> [--agent <id>] [--last N] [--before <event_id>] [--around <event_id>] [--window N] [--topic <text>] [--json]
conversation-esaa decide "<text>" --workspace <path> [--rationale <text>] [--agent <id>] [--source <event_id>]
conversation-esaa task create "<title>" --workspace <path>
conversation-esaa task update <task_id> --workspace <path> [--status <status>] [--next-step <text>]
conversation-esaa task close <task_id> --workspace <path> [--evidence <text>]
\end{lstlisting}

\subsection{\texttt{enable-hooks}}

Manual hook installation is fragile. The v1.1 release therefore includes \texttt{enable-hooks}. For Grok, the command prepares configuration and may register the workspace in a trust list when the local surface permits it. For Claude Code, bootstrap generates the expected configuration and approval may depend on the harness. For Codex, which does not expose an equivalent native hook in this design, the command prepares the watcher.

\begin{lstlisting}
conversation-esaa enable-hooks --agent grok --workspace C:\project --trust
conversation-esaa enable-hooks --agent claude --workspace C:\project --trust
conversation-esaa enable-hooks --agent codex --workspace C:\project --watcher
\end{lstlisting}

The principle is that installation and trust should be operable by command, not by manual editing.

\subsection{\texttt{context}}

The \texttt{context} command turns \texttt{activity.jsonl} into navigable memory. The agent does not need to ingest the whole log; it asks for a window:

\begin{lstlisting}
conversation-esaa context --agent grok --last 20 --workspace C:\project
conversation-esaa context --topic "authentication" --last 5 --workspace C:\project
conversation-esaa context --around evt_123 --window 10 --workspace C:\project
conversation-esaa context --before evt_120 --last 50 --workspace C:\project
\end{lstlisting}

This paginated reading policy reduces the conflict between long horizons and token economy:

\begin{lstlisting}
the full log continues to exist
the agent does not read the full log
the agent reads a projected, filtered, or paginated window
\end{lstlisting}

\subsection{\texttt{decide} and \texttt{task}}

Decisions and tasks are curation, not automatic capture. An agent can record a decision:

\begin{lstlisting}
conversation-esaa decide "Use the hook payload as the primary source" \
  --rationale "Reduces coupling to private vendor formats" \
  --source evt_001 \
  --workspace C:\project
\end{lstlisting}

It can also record operational continuity:

\begin{lstlisting}
conversation-esaa task create "Implement agent-filtered context" --workspace C:\project
conversation-esaa task update CONV-001 --status doing --next-step "Add test" --workspace C:\project
conversation-esaa task close CONV-001 --evidence "conv-test passed" --workspace C:\project
\end{lstlisting}

These commands append events to the log, project read models, and run verification.

\section{Public v1.1.0 Distribution}

The public release was prepared as a greenfield package. This means the published repository contains code, documentation, tests, and synthetic fixtures, but not the private conversational history of the development laboratory. The package's \texttt{activity.jsonl} starts empty after bootstrap.

The public package includes:

\begin{itemize}[leftmargin=*]
  \item \texttt{conversation-esaa.ps1}: the public CLI wrapper;
  \item \texttt{conv-sync.ps1}: the synchronization, projection, and verification engine;
  \item \texttt{conv-bootstrap.ps1}: workspace initialization;
  \item hooks and watchers for supported agents;
  \item \texttt{README.md}, \texttt{PRIVACY.md}, and \texttt{RELEASE.md};
  \item tests \texttt{conv-test.ps1} and \texttt{conv-test-battery.ps1};
  \item a \texttt{.gitignore} excluding generated logs and read models.
\end{itemize}

The v1.1.0 release documents 51 tests in the main battery and validates the public workflow without depending on private history. The initial public commit is \texttt{28aa44a}. This distinction between laboratory and public package is part of the privacy model: development may generate sensitive logs; distribution should start clean.

\section{Privacy}

ESAA-Conversational records the literal text of visible conversation. This is necessary for operational auditability and context recovery, but it creates leakage risk if the directory is versioned carelessly. The v1.1 release adopts four measures:

\begin{enumerate}[leftmargin=*]
  \item the public package is greenfield and carries no private history;
  \item \texttt{.gitignore} excludes \texttt{activity.jsonl}, read models, and generated operational files;
  \item the synchronizer avoids hidden reasoning, raw tool outputs, system prompts, and internal sidechains;
  \item \texttt{PRIVACY.md} states that \texttt{.conversation-esaa/} should be treated as sensitive data.
\end{enumerate}

The v1.1 policy is conservative: public redaction should occur through export or clean bootstrap, not through silent mutation of the local log. Future versions may add \texttt{redact}, redacted export, and formal anonymization trails.

\section{Self-Referential Case Study}

The ESAA-Conversational system was designed, implemented, reviewed, and iteratively corrected inside the very development laboratory whose activity it records. On 21 June 2026, the laboratory \texttt{activity.jsonl} contained 570 events, of which 562 were of type \texttt{conversation\_turn}. The agent distribution in the snapshot was: Codex, 304 events; Claude, 79; Grok, 67; with the remainder consisting of older or unattributed turns without \texttt{agent\_id}.

This case study is self-referential in a strong sense. First, the tool was used to record and project its own genesis: decisions about hook design, command semantics, read-model structure, workspace isolation, privacy policy, and the public distribution strategy all appear as curated events in the same log that the system later projects. Second, heterogeneous agents collaborated on the design and review of ESAA-Conversational without any direct agent-to-agent channel. An agent could be given a focused, deterministic view of another agent's recent work simply by issuing:

\begin{lstlisting}
conversation-esaa context --agent grok --last 20 --workspace C:\project
conversation-esaa context --agent codex --topic "legacy events" --last 10 --workspace C:\project
\end{lstlisting}

This pattern was used repeatedly during development. For example, when refining the \texttt{context} command's handling of legacy events, turns lacking \texttt{event\_id} or \texttt{agent\_id}, Codex was asked to review only the relevant recent Grok iterations. The resulting diagnosis and proposed fix were then recorded via \texttt{decide} and subsequently referenced by other agents. The same mechanism allowed Claude to inspect the evolving task backlog and Grok to verify the projected \texttt{handoff.md} and \texttt{decisions.md} before implementation proceeded.

One concrete defect discovered through this workflow was incomplete filtering logic for legacy events when \texttt{context --topic} was executed against the real laboratory log. The correction was implemented, tested, and documented as a closed task, reinforcing a key design lesson: the public distribution must remain greenfield and strict, while the internal laboratory harness may retain compatibility shims for historical data.

By the end of the recorded period, the log also contained explicit decision and task events created by the agents, demonstrating that the curation surface was actively used rather than remaining theoretical. The entire development trajectory, from initial architecture sketches to the public v1.1.0 release artifacts, tests, documentation, and this paper, therefore constitutes both a validation of the approach and a non-trivial multi-agent workload executed under the governance of the conversational memory layer being described.

The case study provides evidence at two levels: (i) operational viability of the v1.1 design in sustained real-world use, and (ii) the practical value of selective, deterministic handoff across heterogeneous agents without requiring them to share a common runtime or ingest full conversational histories.

\section{Discussion}

\subsection{Token Economy}

The phrase ``zero tokens'' should be understood precisely. Mechanical capture, deduplication, projection, and verification do not use LLM inference. Agents still use tokens when they read context or decide to record a decision. The gain is to remove mechanical history copying from the LLM and replace it with deterministic files and commands.

\subsection{Operational Versus Forensic Auditability}

The v1.1 release provides operational auditability: one can determine which turns were captured, which decisions were recorded, which tasks were opened or closed, and which read model was projected. It does not promise strong forensic auditability. Hash chains, signatures, external anchoring, and formal cold replay belong to a future version if the domain justifies that cost.

\subsection{Why Not Use ESAA Core Directly}

ESAA core remains the reference architecture for agentic governance, but its primary vocabulary concerns work over a project~\cite{esaa}. PARCER is likewise relevant as an output-governance contract for individual agent responses rather than as a shared conversational memory layer~\cite{parcer}. ESAA-Conversational needs commands and events that express memory: synchronizing turns, reading context, recording decisions, preparing handoff, and projecting conversational tasks. Reusing core directly would require mapping conversation turns to artificial tasks or issues, obscuring the domain. A future integration is plausible, but as a backend or specialized profile, not as a replacement for the conversational vocabulary.

\subsection{Limitations}

The v1.1 release, while functional, carries several important limitations that should be taken into account.

\subsubsection*{Technical and Platform Constraints}
\begin{itemize}[leftmargin=*]
  \item The implementation is strictly local, Windows-oriented, and tightly coupled to PowerShell 7. No cross-platform runtime is provided.
  \item Synchronization depends on the log formats and hook surfaces exposed by third-party agents. These interfaces are not under our control and may change without notice, potentially breaking the \texttt{sync} pipeline.
  \item Topic-based retrieval, \texttt{context --topic}, is purely textual and deterministic. No semantic indexing, embeddings, or vector search is implemented.
\end{itemize}

\subsubsection*{Capture and Agent State}
\begin{itemize}[leftmargin=*]
  \item Only visible conversation text is captured. Internal agent state, chain-of-thought reasoning, tool calls, and hidden context are not recorded.
  \item Codex does not expose native hooks in the current design; it relies on a dedicated watcher process, unlike Grok and Claude Code.
  \item Legacy events created before the introduction of \texttt{event\_id}, \texttt{agent\_id}, and \texttt{workspace\_root} fields required backward-compatibility logic. This increases complexity in the projection and context commands.
\end{itemize}

\subsubsection*{Security, Auditability, and Privacy}
\begin{itemize}[leftmargin=*]
  \item The system provides operational auditability but lacks strong forensic guarantees. There are no hash chains, cryptographic signatures, or external anchoring of the event log.
  \item Privacy protection depends on operational discipline. Although the public distribution is greenfield and \texttt{.gitignore} excludes sensitive files, a user may still accidentally commit or share \texttt{activity.jsonl} containing private conversational content.
  \item No redaction or anonymization tooling is implemented. Selective export or automated removal of sensitive turns must be performed manually.
\end{itemize}

\subsubsection*{Concurrency and Operational Performance}
\begin{itemize}[leftmargin=*]
  \item Write concurrency is controlled by a simple lockfile, \texttt{run/*.lock}. This is less robust than a proper single-writer queue or transactional log.
  \item There is no snapshot mechanism. Old history cannot be compacted, forcing agents to potentially read long sequences of cold events when using broad context windows.
  \item Cold replay and deterministic reconstruction of all read models from a clean log, with hash verification, are not supported.
\end{itemize}

\subsubsection*{Validation Scope}
\begin{itemize}[leftmargin=*]
  \item The reported case study, although intensive at 570 events, covers only a single development workspace and a limited set of agents, primarily Codex, Claude, and Grok. Generalization to other environments, larger teams, or different agent combinations remains untested.
\end{itemize}

\section{Future Work}

Natural extensions include:

\begin{enumerate}[leftmargin=*]
  \item \textbf{Snapshots}: consolidate old history into durable state to reduce cold-event reading.
  \item \textbf{Cold replay}: reconstruct all read models from a clean log and compare hashes.
  \item \textbf{Redaction}: generate redacted exports or apply explicit privacy policies.
  \item \textbf{Forensic auditability}: add hash chains, signatures, or external anchoring.
  \item \textbf{Semantic indexes}: add embeddings as derived projections while keeping the event store as the source of truth.
  \item \textbf{Portability}: expand support beyond Windows and PowerShell.
  \item \textbf{ESAA core profile}: investigate whether the ESAA runtime can host the conversational vocabulary without distorting the domain.
\end{enumerate}

\section{Conclusion}

ESAA-Conversational turns continuity across agents into an event-sourcing problem. Instead of relying on manual context copying, the method records visible turns in a shared log, projects compact read models, and offers paginated reading commands. The public v1.1 release confirms the operational viability of the design in a simple local tool: thin hooks, a single CLI, \texttt{activity.jsonl} as source of truth, reconstructible read models, explicit curation, and privacy through greenfield distribution.

The main contribution is not a new semantic memory system, but an architectural boundary: mechanical capture is automatic; curation is judgment; reading is selective; and the complete log remains available without being ingested wholesale by every new agent.

\end{document}